# Factors controlling oxygen interstitial diffusion in the Ruddlesden-Popper oxide $La_{2-x}Sr_xNiO_{4+\delta}$

Shenzhen Xu[1], Ryan Jacobs[1], Dane Morgan[1]

[1]*Department of Materials Science and Engineering, University of Wisconsin – Madison, Madison, WI, 53706, USA*

**Abstract**

The development of Ruddlesden-Popper oxides as oxygen exchange and transport materials for applications such as solid oxide fuel cells, oxygen separation membranes, and chemical looping will benefit from detailed mechanistic understanding of how oxygen is transported through these materials. Using Density Functional Theory, we found there are two distinct oxygen interstitial diffusion mechanisms involving two different oxygen interstitial species that can be active in $La_{2-x}Sr_xNiO_{4+\delta}$, and, we believe, in hyperstoichiometric Ruddlesden-Popper oxides in general. The first mechanism is the previously proposed interstitialcy-mediated mechanism, which consists of diffusing oxide interstitials. The second mechanism is newly discovered in this work, and consists of both oxide and peroxide interstitial diffusing species. This mechanism exhibits a similar or possibly lower migration barrier than the interstitialcy mechanism for high oxidation states. Which mechanism contributes to the oxygen interstitial diffusion is the result of the change in relative stability between the oxide interstitial (2- charge) and peroxide interstitial (1- charge), which directly affects the migration barriers for these two different mechanisms. The stability of the oxide and peroxide, and therefore the competition between the two oxygen diffusion mechanisms, is highly sensitive to the overall oxidation state of the system. Therefore, the oxygen diffusion mechanism is a function of the material composition, oxygen off-stoichiometry, operating temperature and oxygen partial pressure. We also examined the effect of epitaxial strain on both oxygen diffusion mechanisms, and found that tensile and compressive epitaxial strain of up to 2% had less than 100 meV/(% strain) effects on oxygen interstitial formation, migration, and activation energies, and that total achievable activation energy reductions are likely less than 100 meV for up to ±2% epitaxial strain. The presented understanding of factors governing interstitial oxygen diffusion potentially has significant implications for the engineering of Ruddlesden-Popper oxides in numerous alternative energy technologies.

**Table of Contents Figure**

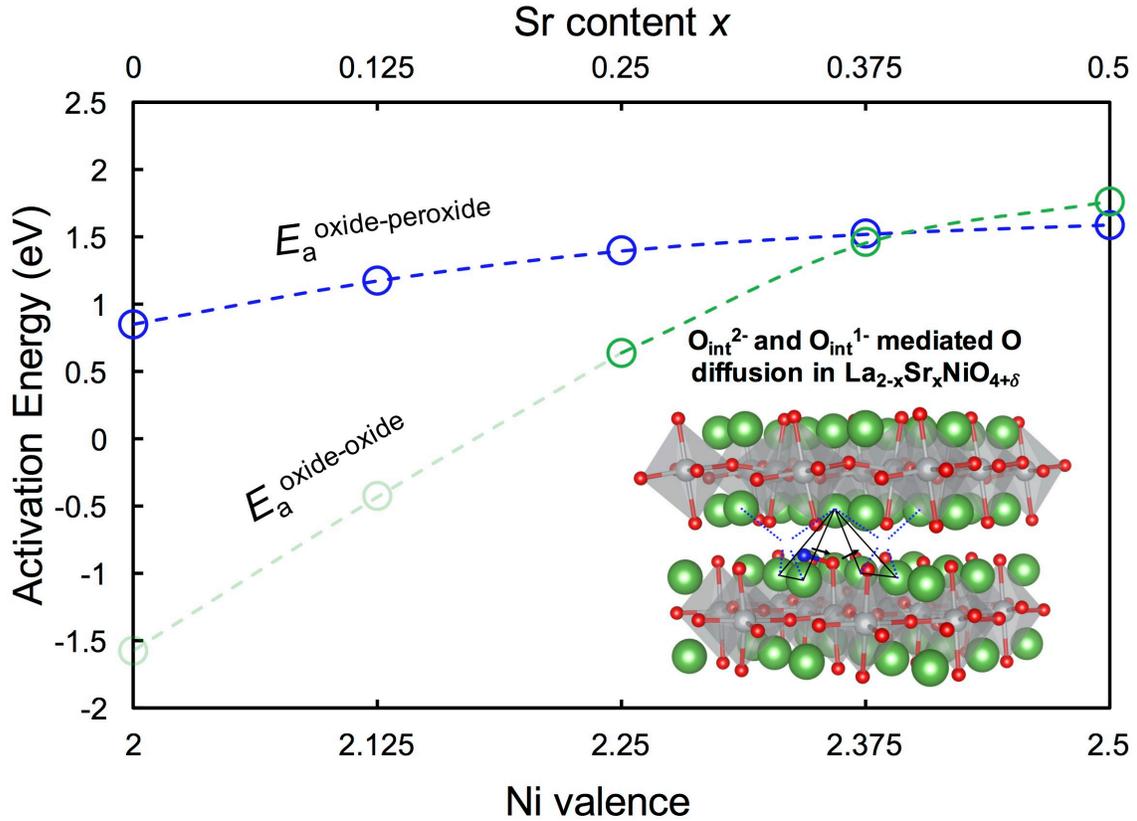

1. **Introduction**

Alternative energy applications such as solid oxide fuel cells (SOFCs), oxygen separation membranes for syngas production and $CO_2$ capture and storage, and chemical looping devices all rely on oxygen active materials that can quickly and efficiently exchange and transport oxygen.[1-6] While transition metal-based perovskite oxides have received widespread attention for their use in these applications as alternatives to rare and expensive precious metal or oxide materials, perovskite-related $n = 1$ Ruddlesden-Popper oxides (chemical formula $(A_{1-x}A'_x)_2BO_{4\pm\delta}$)[7] are also very promising. In particular, over the past 15 years, researchers have demonstrated that the doped lanthanum Ni-, Co-, Cu-, and Fe-based Ruddlesden-Popper materials all show either fast oxygen surface exchange, efficient bulk oxygen transport, or both.[2, 8-19]

A special aspect of Ruddlesden-Popper oxides that sets them apart from perovskites and most other oxygen conductors is that, whereas perovskites and other oxygen conductors tend to either be stoichiometric or contain oxygen vacancies, Ruddlesden-Popper materials may be stoichiometric, or contain some mixture of vacancies and/or interstitials. Based on the material composition and operating environment (temperature and oxygen partial pressure), Ruddlesden-Popper oxides may therefore range from hypostoichiometric (more oxygen vacancies than interstitials) to hyperstoichiometric (more interstitials than vacancies). This flexibility in oxygen stoichiometry results in a rich defect chemistry,[20-24] which in turn enables multiple mechanisms of oxygen transport. In particular, Ruddlesden-Popper materials have been shown to primarily transport oxygen via interstitial- and interstitialcy-mediated diffusion mechanisms, while vacancies appear to play a minor role in diffusion, even for hypostoichiometric materials such as $La_{2-x}Sr_xCuO_{4-\delta}$.[11, 18, 19, 21, 23-27] In addition, there is a high degree of anisotropy in the oxygen transport, where oxygen interstitial diffusion in the rocksalt *a-b* plane is generally at least an order of magnitude faster than along the *c* direction.[28-33] A further unusual property of Ruddlesden-Popper materials is that they can contain interstitials in both oxide and peroxide states,[20, 22, 34-36] and both may potentially participate in diffusion.

The main focus of this paper is to gain an enhanced mechanistic understanding of oxygen interstitial diffusion in hyperstoichiometric Ruddlesden-Popper oxides, using $La_{2-x}Sr_xNiO_{4+\delta}$ as a case-study material. Oxygen interstitial diffusion in Ruddlesden-Popper oxides depends on numerous properties, and in this work we explore the role of three experimentally controllable properties that, when changed, could significantly influence the defect chemistry and oxygen transport properties of the material. These three experimentally controllable properties are material composition, environmental conditions (temperature and oxygen pressure), and strain. Not surprisingly, changes in transition metal element and/or aliovalent doping,[20, 24, 25, 27, 37] and changes in temperature and oxygen pressure,[38-40] will change the defect formation energetics and oxygen surface exchange and migration rates. Strain effects are not well understood, but epitaxial strain has been recently investigated as a property tuning knob for oxygen surface exchange and diffusion in perovskites,[41-44] and has been shown to have a significant impact on the defect concentrations and surface exchange rates in Ruddlesden-Popper oxides.[45-47] Overall, understanding how properties such as composition, operating conditions and strain affect oxygen transport is key for enabling the use of Ruddlesden-Popper materials in an array of alternative energy technologies.

In this study, we used Density Functional Theory (DFT) to investigate the interplay of oxidation state (via compositional change of Sr content $x$ in $La_{2-x}Sr_xNiO_{4+\delta}$) and strain on the oxygen interstitial formation, migration, and resultant diffusion mechanism characteristics, using $La_{2-x}Sr_xNiO_{4+\delta}$ as a case-study Ruddlesden-Popper material. We used Ni valence as a measure of the oxidation state. The Ni valence can be increased by both Sr doping and excess oxygen in the form of oxide interstitials (related to the value of the degree of nonstoichiometry $\delta$). In this work, we will consider $La_{2-x}Sr_xNiO_{4+\delta}$ only with very dilute interstitial oxygen, i.e., $\delta \approx 0$, so that the Ni valence = $2+x+2\delta \approx 2+x$ ($\delta \approx 0$), where $x$ is the Sr content in $La_{2-x}Sr_xNiO_{4+\delta}$. It is worth noting that the trends we observed in this work can be rationalized by considering Ni valence, and we therefore expect these trends to be qualitatively transferable to any $La_{2-x}Sr_xNiO_{4+\delta}$ system regardless of the origin of oxidation (that is, whether the oxidation is from Sr doping, oxide interstitials, or a combination of both). The possible influence of higher $\delta$ values on the barriers and stabilities determined in this work is an important topic but beyond the scope of this initial study. Under the elevated temperatures and near-atmospheric oxygen partial pressure conditions commonly used in SOFCs, oxygen separation and chemical looping applications, the majority defect in Ni-, Co-, Cu-, and Fe-containing Ruddlesden-Popper materials are oxygen interstitials.[8, 11, 34, 48, 49] Thus, in this study we have focused on hyperstoichiometric $La_{2-x}Sr_xNiO_{4+\delta}$. We have found there are two distinct oxygen interstitial diffusion mechanisms involving two different oxygen interstitial species that can be active in $La_{2-x}Sr_xNiO_{4+\delta}$ (**Section 2.1**), and, we believe, in hyperstoichiometric Ruddlesden-Popper oxides in general. In this work, we examined the effect of Ni valence on the oxygen interstitial formation energies (**Section 2.2**), migration barriers (**Section 2.3**), and the resulting activation energies (sum of formation and migration energies, **Section 2.4**) that control which mechanism contributes to oxygen diffusion at a particular Ni valence. Finally, we examined the possible role that modest epitaxial strain may play in changing the overall migration barriers and tuning the Ni valence at which a particular oxygen diffusion mechanism is competitive (**Section 2.5**). Overall, the findings of this work, particularly the understanding of there being two distinct oxygen interstitial diffusion mechanisms involving two different oxygen interstitial species present at different Ni valence states, provide a useful foundation for understanding and engineering of $La_{2-x}Sr_xNiO_{4+\delta}$ and related Ruddlesden-Popper oxides for fast exchange and transport of oxygen.

## 2. Results

### 2.1. Two types of oxygen interstitials and two interstitial diffusion mechanisms

Oxygen interstitials intercalated into a Ruddlesden-Popper material may reside in one of two distinct interstitial states.[20, 22, 34-36] The first state is the oxide state, henceforth denoted as "$O_{int}^{2-}$". **Figure 1** shows that the $O_{int}^{2-}$ resides in the rocksalt layer and is located at the center of the $La_4$ tetrahedron. The second state is the peroxide state, henceforth denoted as "$O_{int}^{1-}$". **Figure 2** shows that $O_{int}^{1-}$ is bonded to one of the apical oxygen atoms in the rocksalt layer, forming an $O_2^{2-}$ dumbbell, with an O-O dumbbell bond length of about 1.5 Å. These two oxygen interstitial species can either be stable or metastable depending on the oxidation state of $La_{2-x}Sr_xNiO_{4+\delta}$. Details of the relative stabilities of these two interstitial species are discussed in **Section 2.2**.

In addition to there being two different oxygen interstitial species, we demonstrate in this work that there are also two distinct oxygen interstitial diffusion mechanisms within the *a-b* plane. In this study, we have focused only on oxygen migration in the *a-b* plane, where oxygen interstitial diffusion is known to be at least an order of magnitude faster than along the *c* direction.[28-32] As discussed in **Section 1**, we only focus on these two distinct interstitial migration mechanisms, and do not consider any vacancy migration mechanisms. The first diffusion mechanism we consider only involves $O_{int}^{2-}$ species, and will henceforth be referred to as the "oxide-oxide" diffusion mechanism. This oxide-oxide pathway is the previously studied interstitialcy-mediated diffusion mechanism,[21, 25, 26, 28] first identified by Chroneos, et al.[28] A schematic plot of the energy landscape and atomistic configurations of the states comprising the oxide-oxide diffusion mechanism are shown in **Figure 1**. A general description of the oxide-oxide diffusion pathway is: {$O_{int}^{2-}$ (stable) (A) → $O_{int}^{2-}$ (transition state) (B) → $O_{int}^{2-}$ (stable) (C)}, where the letters denote the corresponding structures depicted in **Figure 1**. In the oxide-oxide mechanism transition state, the $O_{int}^{2-}$ atom "kicks out" the apical oxygen bound to a Ni atom, leading to the temporary formation of an oxygen vacancy located approximately in the middle of the two nearby $O_{int}^{2-}$ atoms (marked as a red dashed circle in image (B) in **Figure 1**).

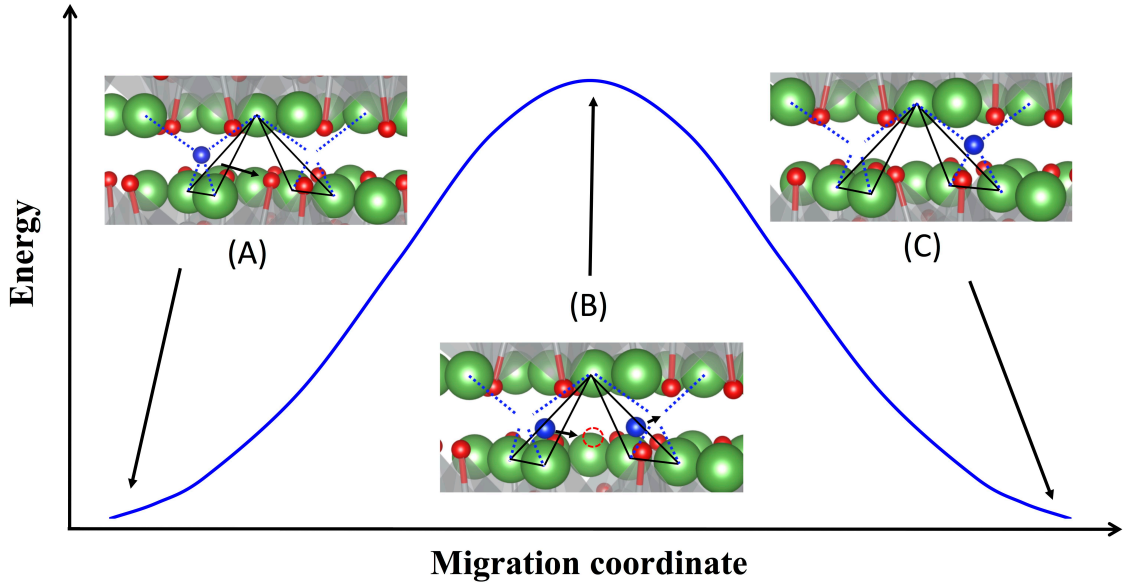

**Figure 1.** The diffusion pathway and the schematic energy landscape of the oxide-oxide interstitialcy-mediated diffusion mechanism. Image (A) shows the atomic configuration of the initial stable $O_{int}^{2-}$ state. Image (C) shows the atomic configuration of the final stable $O_{int}^{2-}$ state. Image (B) is the transition state. The dashed blue lines show the tetrahedral coordination of the $O_{int}^{2-}$, and the dashed red circle in (B) indicates the vacant apical oxygen position. The solid black lines and arrows are meant as guides for the eye of the oxygen interstitial migration path. The blue curve is the schematic energy landscape. The green, silver, and red spheres denote La, Ni, and O atoms, respectively. The blue spheres denote the migrating O atoms.

The second diffusion mechanism is derived by extending the oxide-oxide diffusion model to consider motion of the peroxide interstitial. In particular, the activated state of the oxide-oxide mechanism involves the close approach of an $O_{int}^{2-}$ that kicks out the apical oxygen as shown in **Figure 1B**, creating a transition state with two $O_{int}^{2-}$ species. Exploring this transition state further yielded a new mechanism, where the transition state does not create the apical oxygen vacancy, but instead binds to the apical oxygen to create a stretched peroxide dumbbell with an O-O bond length of about 1.7 Å.

This stretched peroxide dumbbell then relaxes to a shortened bond length of about 1.5 Å to form the $O_{int}^{1-}$ state. This new mechanism involves both the $O_{int}^{2-}$ and $O_{int}^{1-}$ defect species, and will henceforth be referred to as the "oxide-peroxide" diffusion mechanism. A general description of the oxide-peroxide diffusion pathway is {$O_{int}^{2-}$ ((meta)stable) (A) → $O_{int}^{1-}$ ((meta)stable) (B) → $O_{int}^{1-}$ ((meta)stable) (C) → $O_{int}^{2-}$ ((meta)stable) (D)}, and a schematic plot of the corresponding energy landscape and atomic configurations of the (meta)stable states and the transition states are shown in **Figure 2**. In the oxide-peroxide mechanism, the $O_{int}^{2-}$ state can either be energetically stable or metastable compared to the $O_{int}^{1-}$ state, but both are always local minima along the diffusion pathway. In this work, the reported values for the oxide-peroxide migration barriers are always the higher (that is, the rate-limiting) value of migrating from (A) to (B) or (B) to (A) (or equivalently, from (C) to (D) or (D) to (C) due to symmetry) in **Figure 2**. For example, if the $O_{int}^{2-}$ state is more stable than the $O_{int}^{1-}$ state (as shown in **Figure 2**), the rate-limiting diffusion barrier is the energy difference between the transition state and the (stable) $O_{int}^{2-}$ defect state. In **Figure 2**, there is a second, small barrier between the two $O_{int}^{1-}$ peroxide states (image (B) and image (C) in **Figure 2**). Our calculations have shown that the barrier of this oxygen dumbbell kick-out mechanism is less than 100 meV, which is much less than the overall oxide-peroxide diffusion mechanism barrier, and therefore this oxygen dumbbell kick-out process is never predicted to be the rate-limiting step in the cases we have studied. There is no vacancy created in the oxide-peroxide diffusion pathway, and instead the migration oxygens form an $O_2$ dumbbell during the kick-out step. To our knowledge, this particular oxide-peroxide oxygen diffusion mechanism has not been previously proposed in the Ruddlesden-Popper phases. Thus, this study is the first to shed light on the mechanistic aspects of this new oxide-peroxide diffusion mechanism utilizing both peroxide and oxide interstitial species.

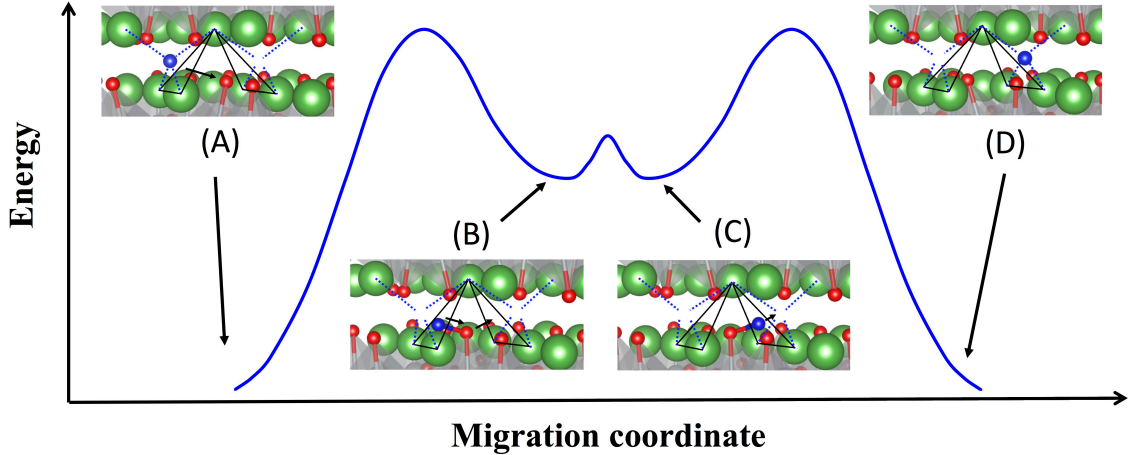

**Figure 2.** The diffusion pathway and the schematic energy landscape of the oxide-peroxide diffusion mechanism. Images (A) and (D) show the atomic configurations of the initial and the final stable $O_{int}^{2-}$ states. Images (B) and (C) show the atomic configurations of the metastable $O_{int}^{1-}$ states. A dumbbell oxygen pair kick-out mechanism occurs between images (B) and (C). The dashed blue lines show the tetrahedral coordination of the $O_{int}^{2-}$. The blue curve is the schematic energy landscape. The green, silver, and red spheres denote La, Ni, and O atoms, respectively. The blue spheres denote the migrating O atoms.

## 2.2. Defect formation energies of the $O_{int}^{2-}$ and $O_{int}^{1-}$ species

In this section, we present our calculated dilute ($\delta \to 0$) defect formation energies for the $O_{int}^{2-}$ and $O_{int}^{1-}$ states as a function of Ni valence for unstrained $La_{2-x}Sr_xNiO_{4+\delta}$. The defect formation energies were calculated assuming an environment corresponding to $T$ = 773 K, $p(O_2)$ = 0.2 atm (these conditions impact the oxygen chemical potential in the gas and vibrational free energy in the solid), which represents a typical operating environment for oxygen active materials used in an array of applications. Additional details on the defect calculations can be found in **Section 5**. From **Figure 3**, we can see that $O_{int}^{2-}$ is the dominant defect species until Ni valence approaches to 2.5 ($x \to 0.5$). $E_f$ ($O_{int}^{2-}$) increases with Ni valence with an overall dramatic energy change (~3 eV) from $Ni^{2+}$ to $Ni^{2.5+}$. By comparison, $E_f$ ($O_{int}^{1-}$) is nearly constant over the entire range of Ni valence. Regarding the trend of $E_f$ ($O_{int}^{2-}$) with Ni valence, the qualitative explanation is that formation of $O_{int}^{2-}$ oxidizes the system, and it becomes progressively more difficult for the system to be oxidized as the Ni valence increases. For a Ni valence of about $Ni^{2.375+}$, there is a clear reduction in slope of $E_f(O_{int}^{2-})$ versus Ni valence. This reduction in slope occurs because Ni is not being oxidized further, and oxygen is instead the species being oxidized.[20] While it can be seen in **Figure 3** that $E_f(O_{int}^{2-})$ and $E_f(O_{int}^{1-})$ are nearly degenerate in the Ni valence range of 2.375-2.5, this degeneracy occurs only briefly with increasing valence as the two energies cross. It is clear from the work of Xie, et al. that as Ni valence continues to increase past 2.5, the difference in $O_{int}^{2-}$ and $O_{int}^{1-}$ formation energies $E_f(O_{int}^{2-})-E_f(O_{int}^{1-})$ continues to increase.[20] Regarding the trend of $E_f(O_{int}^{1-})$ with Ni valence, the $O_{int}^{1-}$ species bonds with an existing O in the lattice, forming an $O_2^{2-}$ dumbbell that does not further oxidize the system. Therefore, the oxidation state of the system (represented by Ni valence) has a minimal impact on $E_f(O_{int}^{1-})$. These results are similar to those previously observed by Xie, et al.[20], and our explanations follow those given in that work. Due to the different dependence of $E_f(O_{int}^{2-})$ and $E_f(O_{int}^{1-})$ on Ni valence, we can see that for $La_{2-x}Sr_xNiO_{4+\delta}$, the relative stability of dilute $O_{int}^{2-}$ versus $O_{int}^{1-}$ switches at about $Ni^{2.4+}$, corresponding to a composition of $x = 0.6$ and $La_{1.4}Sr_{0.6}NiO_{4+\delta}$, assuming $\delta \to 0$. This value of Ni valence is slightly smaller than the results of Xie, et al., who showed the relative stability of $O_{int}^{2-}$ versus $O_{int}^{1-}$ switch at about $x = 0.45$ for $\delta = 0.0625$, which corresponds to $Ni^{2.58+}$.[20] This difference in Ni valence at the crossover is relatively modest and most likely the result of the different $\delta$ values in the present study ($\delta \approx 0$) and those in Xie, et al. ($\delta = 0.0625$) as well as minor differences in the calculation settings and sampled Ni valences in our vs. Xie, et al.'s study.

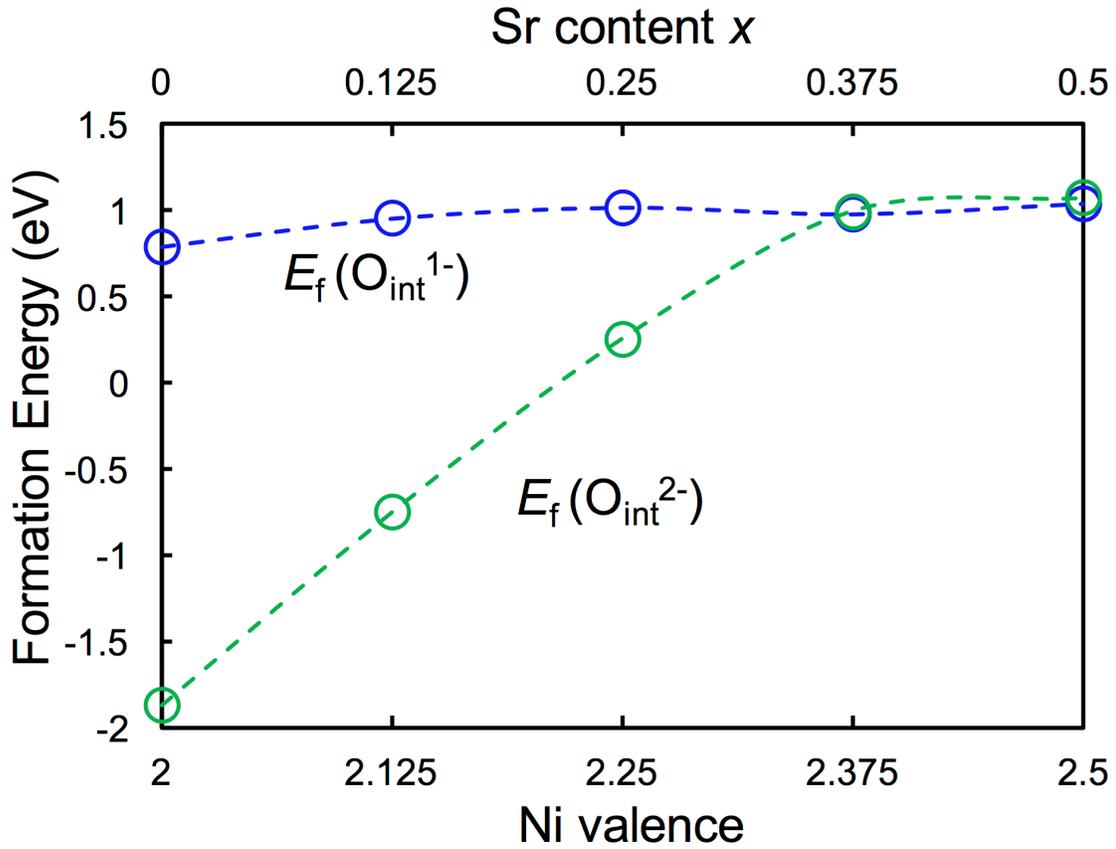

**Figure 3**. Defect formation energies of $O_{int}^{2-}$ oxide state and $O_{int}^{1-}$ peroxide state as a function of Ni valence in $La_{2-x}Sr_xNiO_{4+\delta}$. The green and blue curves correspond to $E_f(O_{int}^{2-})$ and $E_f(O_{int}^{1-})$ states, respectively. The dashed lines serve as a guide for the eye.

## 2.3. Migration barriers of the oxide-oxide and oxide-peroxide mechanisms

**Figure 4** shows the calculated migration barriers $E_b^{oxide\text{-}oxide}$ and $E_b^{oxide\text{-}peroxide}$ for the oxide-oxide and oxide-peroxide diffusion mechanisms, respectively, as a function of Ni valence. For the oxide-oxide diffusion mechanism, $E_b^{oxide\text{-}oxide}$ gradually increases from 0.3 eV to 0.7 eV as the Ni valence increases (more oxidized). The oxide-oxide migration barrier value of 0.3 eV for the case of $x = 0$ compares well with experimental values ranging from 0.19-0.54 eV from isotope exchange and secondary ion mass spectrometry and a value of 0.51 eV obtained from molecular dynamics.[9, 28, 29] The increase in barrier with increasing Ni valence can be understood but examining the hopping mechanism in detail. From analyzing the local structure along the oxide-oxide diffusion pathway, we can see the transition state corresponds to a structure where the $O_{int}^{2-}$ kicks out an apical oxygen atom that is originally bonded to a Ni atom as shown in **Figure 1**. In the transition state shown in **Figure 1**, the diffusing O (shown as blue) O-Ni distance increases from ~2.2Å (bonded to Ni) to ~3.2Å (not bonded to Ni), and the diffusing oxygens do not form a dumbbell pair (the blue O-O distance is 2.7Å, much larger than the peroxide $O_2^{2-}$ bond length of 1.5Å). Therefore, both the diffusing blue O atoms in **Figure 1** are in the 2- state and can be denoted as $O_{int}^{2-}$ species. As the Ni valence increases (more oxidized), more energy is required to remove additional electrons from Ni. In the transition state of the oxide-oxide mechanism there are two oxygen interstitials that have an oxidation state of approximately 2- (i.e., both interstitials are $O_{int}^{2-}$), making it more unfavorable compared to the initial and final states (which each have one $O_{int}^{2-}$) as Ni valence is increased.

For the oxide-peroxide mechanism, $E_b^{oxide-peroxide}$ as a function of Ni valence first decreases until the Ni valence reaches 2.375, then remains approximately flat for 2.375 < Ni valence < 2.5. During this analysis, we found that there is a linear relationship between $E_b^{oxide-peroxide}$ and the energy difference of the $O^{2-}$ and $O^{1-}$ defect states $|E(O_{int}^{2-})-E(O_{int}^{1-})|$. This linear relationship makes physical sense based on the symmetrical diffusion pathway of the oxide-peroxide mechanism and the underlying principle of the Brønsted-Evans-Polanyi relations, which stipulate that the transition state energies of chemical reactions scale linearly with the difference in final and initial state energies (see **Supporting Information (SI) Section 3**, **Figure S1** for more details). This linear scaling relationship is: $E_b^{oxide-peroxide} = 0.831 \times |E(O_{int}^{2-})-E(O_{int}^{1-})|+0.518$ ($R^2 = 0.988$). The behavior of $E_b^{oxide-peroxide}$ as a function of Ni valence can be explained by this linear scaling relation. For 2 < Ni valence < 2.375, $E_b^{oxide-peroxide}$ decreases because $|E(O_{int}^{2-})-E(O_{int}^{1-})|$ decreases, which is due to $E_f(O_{int}^{2-})$ increasing as the system becomes more oxidized. As Ni valence increases, $E_f(O_{int}^{2-})$ is approaching $E_f(O_{int}^{1-})$, so that $|E(O_{int}^{2-})-E(O_{int}^{1-})|$ is approaching zero. For Ni valence > 2.375 the energy difference between $E_f(O_{int}^{1-})$ and $E_f(O_{int}^{2-})$ is nearly unchanged with respect to Ni valence (as shown in **Figure 3**) and therefore $E_b^{oxide-peroxide}$ is approximately constant in this range (as shown in **Figure 4**).

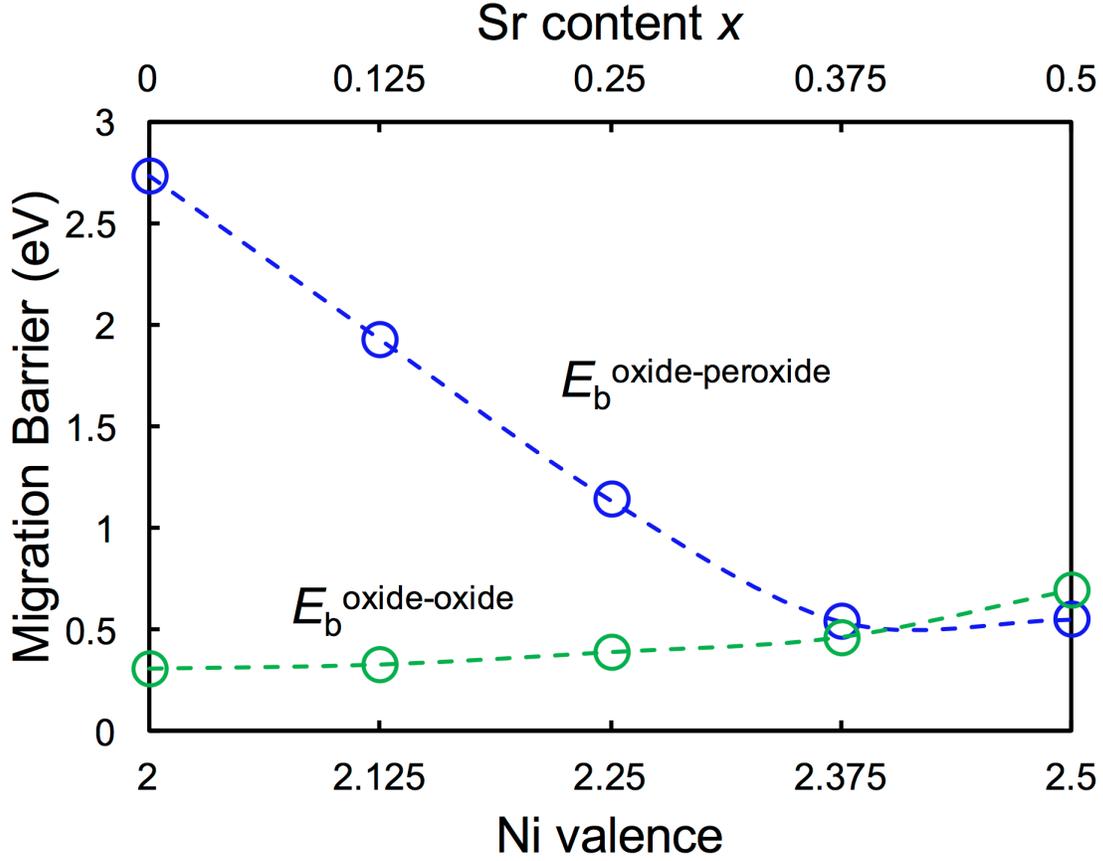

**Figure 4**. Calculated migration barriers of the oxide-oxide and oxide-peroxide diffusion mechanisms as a function of Ni valence in $La_{2-x}Sr_xNiO_{4+\delta}$. The green and blue curves correspond to $E_b^{\text{oxide-oxide}}$ and $E_b^{\text{oxide-peroxide}}$, respectively. The dashed lines serve as a guide for the eye.

## 2.4. Activation energy comparison between oxide-oxide and oxide-peroxide mechanisms

Here, we describe the criterion used in this work to determine which oxygen diffusion mechanism is dominant at a certain Ni valence and strain condition. In general, the chemical diffusion coefficient $D$ of a diffusing species can be written:

$$D \propto e^{\frac{-E_a}{kT}} = C \times e^{\frac{-E_b}{kT}} \propto e^{\frac{-E_f}{kT}} \times e^{\frac{-E_b}{kT}}, \qquad (1)$$

where $C$ is the concentration of the diffusing species (in this case, interstitials) and $E_a = E_f + E_b$ is the effective activation energy for diffusion. The constants of proportionality can vary significantly under different conditions, but they are generally not exponential in temperature and therefore it is reasonable to assume at this level of analysis that relative diffusion rates between different oxygen interstitial migration mechanisms are dominated by $E_a$. This assumes that relative entropic terms do not play a major role in our competing diffusion mechanisms, which is an assumption that needs further study. Based on these assumptions, we evaluate which diffusion mechanism is relevant as a function of Ni valence by calculating the activation energy of the oxide-oxide ($E_a^{\text{oxide-oxide}}$) and oxide-peroxide ($E_a^{\text{oxide-peroxide}}$) mechanisms, where the activation energy for the oxide-oxide mechanism is $E_a^{\text{oxide-oxide}} = E_f(O_{\text{int}}^{2-}) + E_b^{\text{oxide-oxide}}$ and the activation energy of the oxide-peroxide mechanism is $E_a^{\text{oxide-peroxide}} = \min\{E_f(O_{\text{int}}^{2-}), E_f(O_{\text{int}}^{1-})\} + E_b^{\text{oxide-peroxide}}$. As a reminder, for the oxide-oxide mechanism only the $O_{\text{int}}^{2-}$ species is involved in diffusion, so $E_f(O_{\text{int}}^{2-})$ is the only relevant formation energy term. For the oxide-peroxide mechanism, either the $O_{\text{int}}^{2-}$ or the $O_{\text{int}}^{1-}$ may function as the initial state and activated state species based on their relative stability (if formation of $O_{\text{int}}^{2-}$ is favored at a particular Ni valence, then $O_{\text{int}}^{1-}$ is the activated state as shown in **Figure 2**, and vice versa), so the activation energy for this diffusion mechanism requires the minimum value of $E_f(O_{\text{int}}^{2-})$ and $E_f(O_{\text{int}}^{1-})$.

The values of $E_a^{\text{oxide-oxide}}$ and $E_a^{\text{oxide-peroxide}}$ as a function of Ni valence are shown in **Figure 5**. For 2 < Ni valence < 2.25, $E_a^{\text{oxide-oxide}} \ll E_a^{\text{oxide-peroxide}}$ and the oxide-oxide diffusion mechanism is dominant. From **Figure 5**, at low Ni valence the value of $E_a^{\text{oxide-oxide}}$ is negative due to the negative defect formation energy in this regime (see **Figure 3**). In **Figure 5**, these negative $E_a^{\text{oxide-oxide}}$ values are greyed out due to their unphysical nature. They are unphysical as we have calculated $E_a^{\text{oxide-oxide}}$ assuming dilute oxygen diffusion ($\delta \to 0$). However, the negative defect formation energy indicates the equilibrium oxygen interstitial content will be larger than we have simulated here, and will lead to increasing $E_a^{\text{oxide-oxide}}$ due to interactions between defects, which can be significant.[20] As remarked earlier, the possible influence of higher $\delta$ values on our calculated values is an important topic but beyond the scope of this initial study. As the Ni valence continues to increase and approaches 2.5, $E_a^{\text{oxide-oxide}}$ keeps increasing and becomes comparable to (and even larger than) $E_a^{\text{oxide-peroxide}}$ due to the increasing instability of $O_{\text{int}}^{2-}$, and the peroxide-oxide mechanism emerges as a competitive oxygen diffusion pathway. We expect that the qualitative feature of the oxide-peroxide mechanism being an important oxygen diffusion channel continues to hold as the Ni valence increases because higher Ni valence will tend to further destabilize the $O_{int}^{2-}$ formation (as discussed in **Section 2.2)**, making the oxide-oxide mechanism less competitive at even higher Ni valence states. Although our current study cannot predict a highly quantitative Ni valence value where the mechanism-switching point occurs, the qualitative trends of the activation energy profiles with Ni valence as shown in **Figure 5** provide strong evidence for the existence of a previously undiscovered oxide-peroxide diffusion mechanism in La$_{2-x}$Sr$_x$NiO$_{4+\delta}$ at a relatively high Ni valence state which involves both oxide and peroxide interstitial species. This high Ni valence state might occur due to Sr or other aliovalent doping, excess oxygen interstitials, a combination of both, or some other oxidizing mechanism.

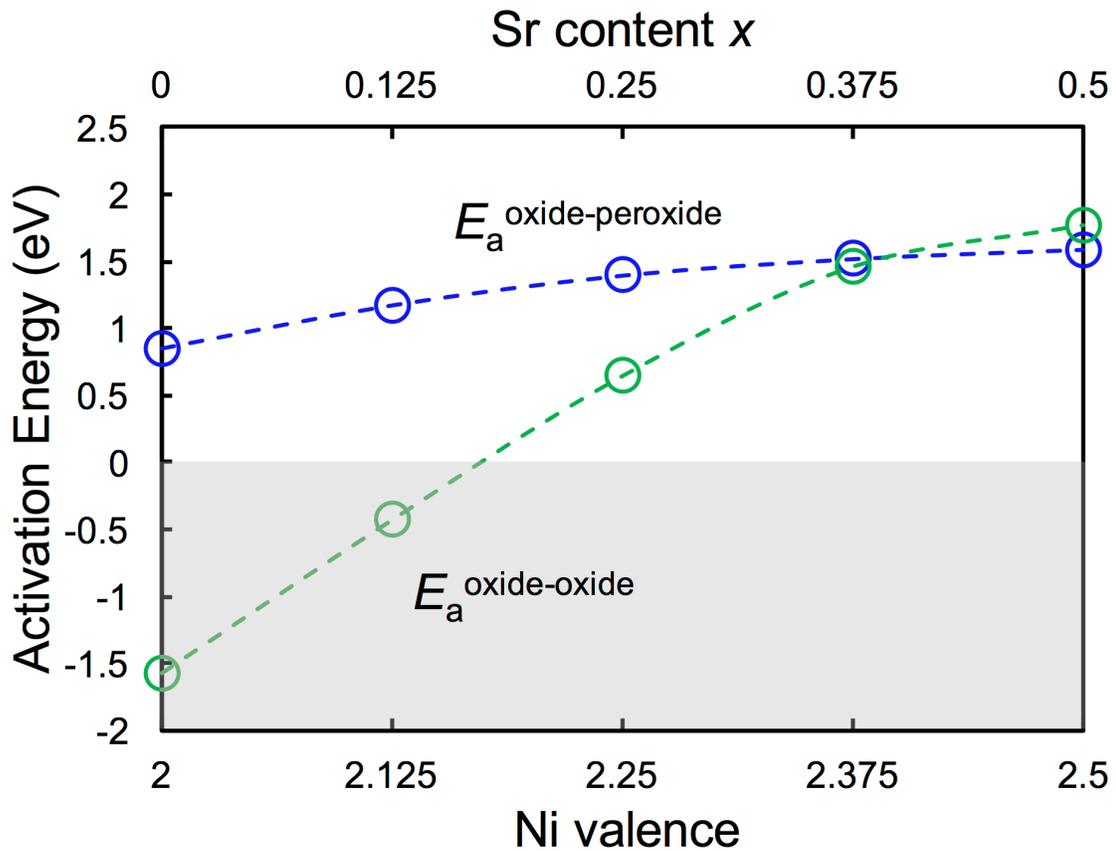

**Figure 5**. Activation energies $E_a^{\text{oxide-oxide}}$ (green) and $E_a^{\text{oxide-peroxide}}$ (blue) of the oxide-oxide and oxide-peroxide diffusion mechanisms, respectively, as a function of Ni valence in La$_{2-x}$Sr$_x$NiO$_{4+\delta}$. The dashed lines serve as a guide for the eye. The negative activation energies are shaded out, and are a consequence of the negative defect formation energies.

## 2.5. Strain effects on $O_{int}^{2-}$ and $O_{int}^{1-}$ formation and migration

As discussed in **Section 1**, applying epitaxial strain in the *a-b* plane of Ruddlesden-Popper materials may change the defect formation and migration energetics. In this section, we have investigated the effect of modest epitaxial strain ranging from 2% tensile strain to 2% compressive strain on the formation and migration energetics of $La_{2-x}Sr_xNiO_{4+\delta}$ as a function of Ni valence.

**Figure 6** contains the formation energies of $O_{int}^{2-}$ and $O_{int}^{1-}$ (**Figure 6A** and **Figure 6B**), migration barriers $E_b^{\text{oxide-oxide}}$ and $E_b^{\text{oxide-peroxide}}$ (**Figure 6C** and **Figure 6D**), and activation energies $E_a^{\text{oxide-oxide}}$ and $E_a^{\text{oxide-peroxide}}$ (**Figure 6E** and **Figure 6F**) as a function of strain for Ni valences of $Ni^{2+}$ (square symbols), $Ni^{2.25+}$ (circle symbols) and $Ni^{2.5+}$ (diamond symbols). In **Figure 6A**, for $O_{int}^{2-}$, tensile (compressive) strain increases (reduces) $E_f(O_{int}^{2-})$ by order 100-150 meV/(% strain). For $O_{int}^{2-}$, the change in $E_f(O_{int}^{2-})$ of about 100-150 meV/(% strain) occurs at relatively low Ni valence ($Ni^{2+} \rightarrow Ni^{2.25+}$), and the magnitude of the $E_f(O_{int}^{2-})$ change is lessened as Ni valence increases. At $Ni^{2.5+}$, the effect of strain on $E_f(O_{int}^{2-})$ switches, with tensile (compressive) strain now resulting in a reduced (increased) $E_f(O_{int}^{2-})$, though the quantitative change is small (<50 meV/(% strain)), so $E_f(O_{int}^{2-})$ is not significantly affected by this level of epitaxial strain at a high Ni valence.

In contrast to $O_{int}^{2-}$, for $O_{int}^{1-}$ tensile (compressive) strain reduces (increases) $E_f(O_{int}^{1-})$ by order 15-75 meV/(% strain). The trend of change in $E_f(O_{int}^{1-})$ with strain as a function of Ni valence is both opposite of $O_{int}^{2-}$ and lower in magnitude. The magnitude of the change in formation energy caused by tensile strain increases from about 15 meV/(% strain) to about 55 meV/(% strain) as Ni valence increases from $Ni^{2+} \rightarrow Ni^{2.5+}$. For compressive strain, however, the corresponding change in formation energy is almost constant at about 75 meV/(% strain) throughout $Ni^{2+} \rightarrow Ni^{2.5+}$.

The origin of the opposite strain response of $O_{int}^{2-}$ versus $O_{int}^{1-}$ has not been rigorously determined here. However, examining the change in lattice constants (specifically the *c/a* ratio) of fully relaxed $La_2NiO_{4+\delta}$ prior and after formation of $O_{int}^{2-}$ and $O_{int}^{1-}$ defects can provide a qualitative explanation. The *c/a* ratio of a fully relaxed $O_{int}^{2-}$ ($O_{int}^{1-}$) supercell is 2.24 (2.21), whereas the *c/a* ratio of the undefected $La_2NiO_4$ material is 2.21. Therefore, the intercalation of $O_{int}^{2-}$ tends to elongate the $La_2NiO_{4+\delta}$ lattice along the *c*-axis, and an epitaxial compressive strain lowers the value of $E_f(O_{int}^{2-})$ as compressive strain in the *a-b* plane results in an elongation of the *c*-axis. For the case of $O_{int}^{1-}$, from examining the geometry of the $La_2NiO_{4+\delta}$ lattice containing an $O_{int}^{1-}$, the $O_2^{2-}$ dumbbell pair is situated within a single La-O plane, and the intercalation of $O_{int}^{1-}$ tends to expand the La-O plane, resulting in an increase of the *a*-axis, leading to a lower *c/a* ratio. Because of this expansion of the *a*-axis, modest epitaxial tensile strain can stabilize the formation of $O_{int}^{1-}$. The observations discussed above of how strain results in changes in defect formation energy as explained by changes in lattice volume are qualitatively consistent with previously observed trends of defect formation in $La_{1.85}Sr_{0.15}CuO_{4\pm\delta}$ as a function of strain via examination of lattice volume changes and defect formation volumes.[45]

In addition to these trends of defect formation energy with strain, we note that the shape of the formation energy versus strain curves in **Figure 6** have curvatures that show weak parabolic behavior ($E_f(O_{int}^{1-})$, **Figure 6B**), or curvatures that are small enough that the behavior appears linear ($E_f(O_{int}^{2-})$, **Figure 6A**). Previous studies on a series of perovskites[50] and Ruddlesden-Popper $La_{1.85}Sr_{0.15}CuO_{4\pm\delta}$[45] show that the formation energies of vacancies (for perovskites) and vacancies and interstitials (for $La_{1.85}Sr_{0.15}CuO_{4\pm\delta}$) as a function of strain are parabolic with curvatures higher than what we find here. We speculate that the lower curvature of interstitials in $La_{2-x}Sr_xNiO_{4+\delta}$ seen here could result from factors such as the redox of Ni behaving differently than Cu in the Ruddlesden-Popper structure with strain, and structural differences either from different bulk symmetries or internal octahedral tilting patterns changing with strain.

From the trends of $E_f(O_{int}^{2-})$ and $E_f(O_{int}^{1-})$ in **Figure 6A** and **Figure 6B**, it is evident that tensile (compressive) strain tends to move the $O_{int}^{2-}/O_{int}^{1-}$ stability switching point toward lower (higher) Ni valence. Overall, our results show that modest epitaxial tensile and compressive strains have a small effect on the magnitude of oxygen interstitial formation energies for a range of Ni valence. These observed small changes in defect formation energy on the order of 100 meV/(% strain) are qualitatively consistent with the small changes in both oxygen vacancy and interstitial formation as a function of strain in $La_{1.85}Sr_{0.15}CuO_{4\pm\delta}$ as demonstrated in the work of Meyer, et al.[45]

**Figure 6C** and **Figure 6D** show the effect of epitaxial strain on the migration barriers $E_b^{\text{oxide-oxide}}$ and $E_b^{\text{oxide-peroxide}}$, respectively. Based on the trends in formation energies with strain discussed above, tensile (compressive) strain reduces (increases) the energy difference $E(O_{\text{int}}^{2-})-E(O_{\text{int}}^{1-})$. Therefore, the value of $E_b^{\text{oxide-peroxide}}$ is expected to be reduced (increased) in tensile (compressive) strain based on the linear scaling relation between $E_b^{\text{oxide-peroxide}}$ and $|E(O_{\text{int}}^{2-})-E(O_{\text{int}}^{1-})|$ discussed in **Section 2.3** and **SI Section 3**. From **Figure 6C** and **Figure 6D**, it is clear that tensile (compressive) strain reduces (increases) both $E_b^{\text{oxide-peroxide}}$ and $E_b^{\text{oxide-oxide}}$, though by different magnitudes. For $E_b^{\text{oxide-oxide}}$, the change caused by epitaxial strain is about 45 meV/(% strain), while the change for $E_b^{\text{oxide-peroxide}}$ is about 100-150 meV/(% strain) for Ni valence ranging from $Ni^{2+}$-$Ni^{2.25+}$ and approaches zero when the system is highly oxidized at $Ni^{2.5+}$. We note that the linear relationship between $E_b^{\text{oxide-peroxide}}$ and $|E(O_{\text{int}}^{2-})-E(O_{\text{int}}^{1-})|$ is found to hold for changes in these quantities with strain as well as overall Ni valence (see **Figure S1** in **SI Section 3**).

**Figure 6E** and **Figure 6F** show the effect of epitaxial strain on the activation energies $E_a^{\text{oxide-oxide}}$ and $E_a^{\text{oxide-peroxide}}$, respectively. For the oxide-oxide mechanism at low Ni valence of $Ni^{2+}$-$Ni^{2.25+}$, tensile (compressive) strain results in an increase (decrease) in $E_a^{\text{oxide-oxide}}$ of about 100 meV/(% strain). As in the case of **Figure 5**, in **Figure 6E** the activation energies for $E_a^{\text{oxide-oxide}}$ at $Ni^{2+}$ are negative, which is again the result of the negative defect formation energy as discussed above. For the oxide-peroxide mechanism at low Ni valence, $E_a^{\text{oxide-peroxide}}$ is nearly invariant as a function of strain. At the higher Ni valence of $Ni^{2.5+}$, both $E_a^{\text{oxide-oxide}}$ and $E_a^{\text{oxide-peroxide}}$ decrease as a function of strain, by about 75 and 50 meV/(% strain), respectively.

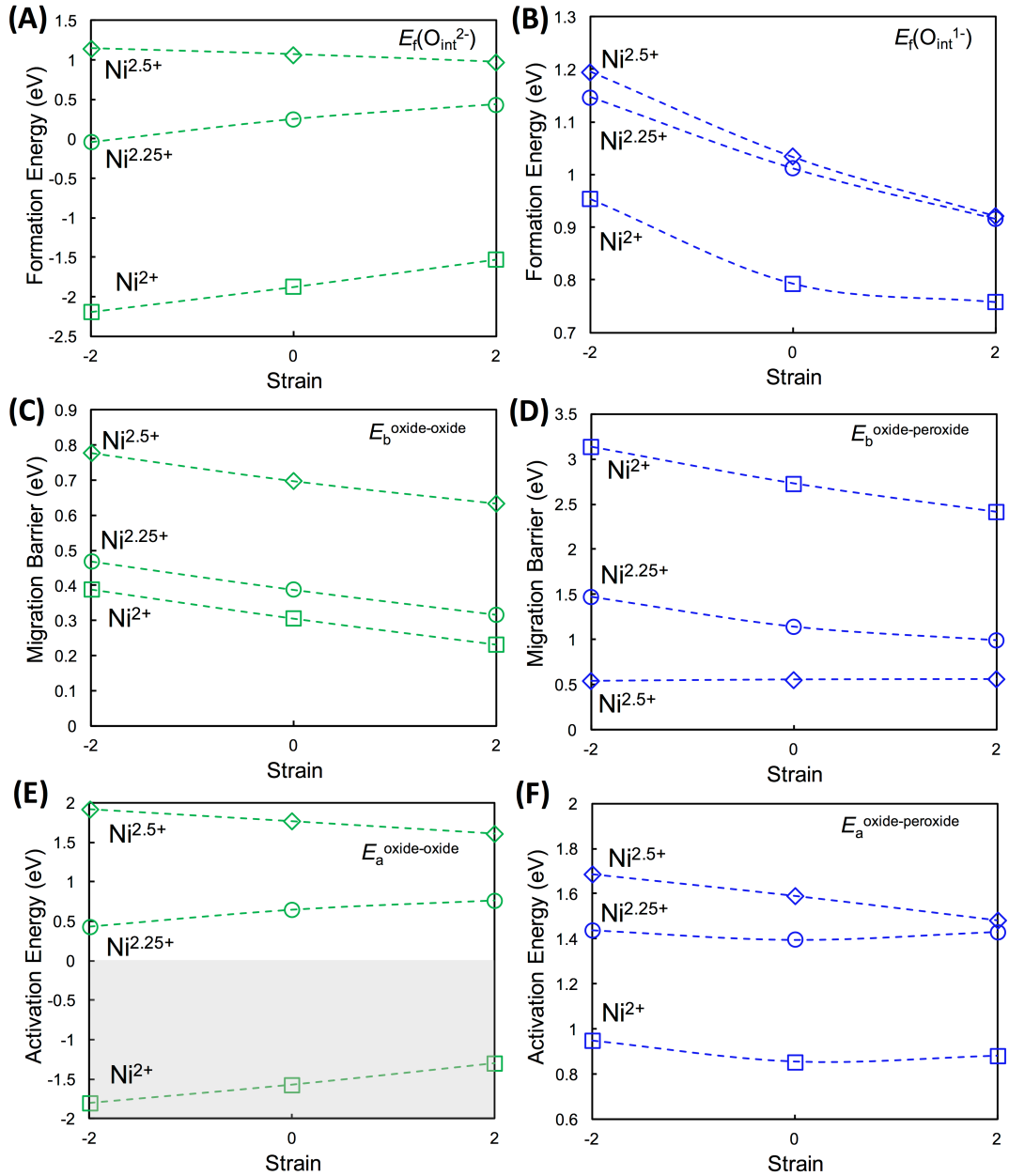

**Figure 6**. Effect of 2% epitaxial tensile and 2% compressive strain on the oxygen interstitial formation energies, migration barriers and activation energies in La$_{2-x}$Sr$_x$NiO$_{4+\delta}$. (A) Response of $E_f(O_{int}^{2-})$ and (B) $E_f(O_{int}^{1-})$ to strain for different Ni valence. (C) Response of $E_b^{oxide-oxide}$ and (D) $E_b^{oxide-peroxide}$ to strain for different Ni valence. (E) Response of $E_a^{oxide-oxide}$ and (F) $E_a^{oxide-peroxide}$ to strain for different Ni valence. For all plots, the square, circle, and diamond symbols correspond to Ni valence of Ni$^{2+}$, Ni$^{2.25+}$, and Ni$^{2.5+}$, respectively. The dashed lines serve as a guide for the eye. The negative activation energies in (E) are shaded out, and are a consequence of the negative defect formation energies.

## 3. Discussion

Our DFT calculations have explicitly shown the existence of a new oxide-peroxide diffusion mechanism in the Ruddlesden-Popper oxide $La_{2-x}Sr_xNiO_{4+\delta}$, which involves both $O_{int}^{2-}$ and $O_{int}^{1-}$ interstitial species and originates from the change in relative stability of $O_{int}^{2-}$ (less stable) to $O_{int}^{1-}$ (more stable) at high Ni valence ($\sim Ni^{2.5+}$, Sr content $x \sim 0.5$ and $\delta \sim 0$). This change in relative stability is due to oxidation of $La_{2-x}Sr_xNiO_{4+\delta}$ via Sr doping, making $O_{int}^{2-}$ formation harder compared to $O_{int}^{1-}$ formation. Thus, the formation energy difference $|E(O_{int}^{2-})-E(O_{int}^{1-})|$ becomes smaller, leading to a lower value of $E_b^{oxide-peroxide}$ due to a more stable activated state. In our case study of the $La_{2-x}Sr_xNiO_{4+\delta}$ system with $\delta \approx 0$, the oxide-peroxide mechanism becomes a competitive oxygen diffusion pathway at a Ni valence of approximately 2.4 (Sr doping level $x = 0.4$ at and $\delta \sim 0$).

In addition to the fundamental understanding of different oxygen interstitial diffusion mechanisms as a function of Ni valence state, our results also indicate that a low activation barrier of less than 1 eV can be achieved at a relatively low Ni valence state of < 2.3 as shown in **Figure 5.** In this low Ni valence regime, the oxide-oxide diffusion mechanism dominates. If the system is highly oxidized such that the Ni valence is increased to about 2.5, $E_a^{oxide-peroxide}$ becomes comparable to $E_a^{oxide-oxide}$ and the oxide-peroxide mechanism thus emerges as a competitive diffusion pathway for $O_{int}$. At this high Ni valence, $E_f(O_{int}^{2-})$ and $E_f(O_{int}^{1-})$ are large, indicating an extremely low oxygen interstitial content if this Ni valence is created by Sr doping, which is consistent with experimental observations from Nakamura, et al.[38] Thus, under this highly oxidized condition, the flux density of diffusing oxygen may be low enough that the material can no longer transport sufficient oxygen to yield desired device performance. Indeed, numerous studies have shown that $La_2NiO_{4+\delta}$ exhibits faster oxygen transport properties than $La_{2-x}Sr_xNiO_{4+\delta}$ with $x > 0$, and that oxygen diffusion as measured through diffusivity, oxygen permeation flux, or calculated as a migration barrier, is decreased as $\delta$ increases.[8, 26, 28, 39] These observed trends from previous studies are consistent with the results and trends depicted in **Figure 5**, where increasing Ni valence yields a more positive activation barrier, thus resulting in slower oxygen transport. Overall, for the purpose of engineering a Ruddlesden-Popper material for application as a fast oxygen diffuser, for the case of $La_{2-x}Sr_xNiO_{4+\delta}$ a low Ni valence (low Sr doping level) can achieve both a high concentration of oxygen interstitials (**Figure 3**) and a low migration barrier with the oxide-oxide mechanism being dominant (**Figure 4**), explaining the observed good oxygen transport properties of this material. The oxide-peroxide mechanism therefore does not appear to be a competitive diffusion mechanism in the doping and oxygen activity domains of interest for the present applications of $La_{2-x}Sr_xNiO_{4+\delta}$, though the oxide-peroxide mechanism may be competitive in other hyperstoichiometric Ruddlesden-Popper oxides under application-relevant conditions.

While we have limited our current study to characterizing oxygen interstitial migration in $La_{2-x}Sr_xNiO_{4+\delta}$ as a case study, we believe the oxide-oxide and oxide-peroxide diffusion mechanisms examined here represent general diffusion mechanisms that may be active in a wide variety of hyperstoichiometric Ruddlesden-Popper oxides. As the cation valence state (that is, the overall oxidation state of the material) is a key quantity to assess which mechanism contributes most to oxygen diffusion, the A-site doping level, B-site transition metal type, and magnitude of oxygen hyperstoichiometry $\delta$ in a particular Ruddlesden-Popper oxide will all impact whether oxygen interstitial diffusion is governed by the interstitialcy-mediated oxide-oxide or our newly discovered oxide-peroxide mechanism, thus potentially impacting device design and operating principles. As discussed above, while interstitial diffusion in $La_{2-x}Sr_xNiO_{4+\delta}$ may not be facilitated by the oxide-peroxide mechanism under the conditions that also promote high oxygen interstitial content, other Ruddlesden-Popper oxides with different material compositions may have interstitial diffusion primarily occurring via the oxide-peroxide mechanism at lower levels of oxidation, resulting in a material that has a high value of $\delta$ and oxygen diffusion governed by this newly discovered oxide-peroxide mechanism.

As discussed above, the presence of lattice strain can also have a significant effect, although likely not very large for high temperatures and strains less than ≈2%. To clarify this further, here we discuss how one may lower the activation energy (both $E_a^{\text{oxide-peroxide}}$ and $E_a^{\text{oxide-oxide}}$), and tune the Ni valence state where the oxide-peroxide mechanism becomes competitive with respect to the oxide-oxide mechanism by applying epitaxial strain. The main reason the oxide-peroxide mechanism is not dominant over a wide range of Ni valence is the high value of $E_f(O_{\text{int}}^{1-})$. Applying tensile strain will lower $E_f(O_{\text{int}}^{1-})$, thus shifting $E_\text{a}^\text{oxide-peroxide}$ down and resulting in the emergence of the oxide-peroxide diffusion mechanism at a lower Ni valence.

However, the overall reduction in activation barrier is quite limited at 2% tensile strain. While we have found this to be true only for our case study of $La_{2-x}Sr_xNiO_{4+\delta}$, it is reasonable to expect that other hyperstoichiometric Ruddlesden-Popper oxides that are chemically similar (for example, $Nd_2NiO_4$, $La_2CoO_4$) will behave similarly. We can thus speculate that epitaxial strain, at least at the modest level of a couple percent, is unlikely to turn poor oxygen diffusing Ruddlesden-Popper materials into good oxygen diffusers. However, this modest level of strain may result in some quantitative enhancement, and this enhancement could result in dramatic improvements for lower-temperature (e.g., room temperature) applications.

## 4. Summary and Conclusions

In this work, we used DFT to study the effect of Ni valence state (which we take to represent overall oxidation state, although it is produced via changing the Sr doping level) and epitaxial strain on the oxygen interstitial formation and migration energetics in Ruddlesden-Popper $La_{2-x}Sr_xNiO_{4+\delta}$. The main findings of this study are:

(1) There are two potential oxygen interstitial diffusion mechanisms in $La_{2-x}Sr_xNiO_{4+\delta}$, and, we believe, hyperstoichiometric Ruddlesden-Popper oxides in general, that are mediated by two different oxygen interstitial species: the oxide interstitial $O_\text{int}^{2-}$ and the peroxide interstitial $O_\text{int}^{1-}$. The first diffusion mechanism is the previously identified "oxide-oxide" mechanism, which is interstitialcy-mediated, where both the initial and the final states of the oxygen interstitial are in $O_\text{int}^{2-}$ states (see **Figure 1**). The second diffusion mechanism is the "oxide-peroxide" mechanism, where both the $O_\text{int}^{2-}$ and $O_\text{int}^{1-}$ species are involved (see **Figure 2**).

(2) Regarding the oxygen interstitial formation energies, $E_\text{f}\,(O_\text{int}^{2-})$ increases as the system becomes more oxidized (that is, as Ni valence increases), while $E_\text{f}\,(O_\text{int}^{1-})$ is almost independent of Ni valence (see **Figure 3**). As $E_\text{f}(O_\text{int}^{2-})$ increases with Ni valence and $E_\text{f}\,(O_\text{int}^{1-})$ is nearly independent of Ni valence, a switch in relative stability of $O_{int}^{2-}$ (less stable) and $O_{int}^{1-}$ (more stable) is expected as Ni valence increases, which we predict occurs at a Ni valence of approximately 2.4+.

(3) Regarding the migration barriers of the oxide-oxide ($E_b^{oxide-oxide}$) and oxide-peroxide ($E_b^{oxide-peroxide}$) mechanisms, our DFT calculations show that $E_b^{oxide-oxide}$ gradually increases (from about 0.3 eV to 0.7 eV) as Ni valence increases from 2 to 2.5, while $E_b^{oxide-peroxide}$ is determined by the formation energy difference between $O_{int}^{2-}$ and $O_{int}^{1-}$. Combining the formation energy and migration barrier results, our study indicates that the oxide-peroxide mechanism becomes competitive with the oxide-oxide mechanism at a Ni valence state above ~2.4+ in $La_{2-x}Sr_xNiO_{4+\delta}$.

(4) Modest epitaxial tensile strain of 2% produces a slight reduction in $E_b^{oxide-oxide}$ and $E_b^{oxide-peroxide}$ over most typical Ni valence values. For $E_b^{oxide-oxide}$, the reduction is 39, 38, and 35 meV/(% strain) for Ni valences of 2, 2.25, and 2.5, respectively. For $E_b^{oxide-peroxide}$, the reduction is about 180, 120, and 0 meV/(% strain) for Ni valences of 2, 2.25 and 2.5, respectively. Although the variation of $E_b^{oxide-peroxide}$ due to epitaxial tensile strain at $Ni^{2+}$ is relatively large, this oxide-peroxide mechanism has a negligible contribution to interstitial diffusion at $Ni^{2+}$ as the difference in activation energy between the oxide-peroxide and oxide-oxide mechanisms at $Ni^{2+}$ is more than 2.5 eV. Therefore, while the change in $E_b^{oxide-peroxide}$ with strain is relatively large at 180 meV/(% strain) at $Ni^{2+}$, no practical amount of strain could be applied to $La_{2-x}Sr_xNiO_{4+\delta}$ to make the oxide-peroxide mechanism competitive at $Ni^{2+}$. Overall, these changes in migration barrier with strain are on the order of 50-100 meV/(% strain), and thus unlikely to result in a dramatic increase in oxygen diffusivity or large change in the oxidation state value where the oxide-peroxide mechanism emerges as a competitive diffusion mechanism, at least at typical device operating temperatures of over 700K.

(5) The new oxide-peroxide oxygen diffusion mechanism predicted in this work has a high activation energy over most values of Ni valence and thus won't contribute to oxygen diffusion in $La_{2-x}Sr_xNiO_{4+\delta}$ unless the system is highly oxidized to a Ni valence above ~2.4. For the specific case of $La_{2-x}Sr_xNiO_{4+\delta}$, a high oxidation state typically results in a low concentration of oxygen interstitials, indicating that this oxide-peroxide mechanism is not expected to be highly active in $La_{2-x}Sr_xNiO_{4+\delta}$. However, we believe this new oxide-peroxide diffusion mechanism may be active in other Ruddlesden-Popper materials, due to the fact that which diffusion mechanism is active will depend on a variety of factors, including the A-site doping, B-site transition metal, oxygen hyperstoichiometry (which is a function of the composition and the equilibration conditions $T$ and $p(O_2)$), and, to a lesser extent, strain.

Overall, this work provides new fundamental understanding of oxygen diffusion in Ruddlesden-Popper oxides. This includes the prediction of a new oxygen diffusion mechanism involving both oxide and peroxide interstitial species, which is expected to occur under conditions where the material is highly oxidized. Mechanistic insight of oxygen transport in Ruddlesden-Popper materials is a key piece of physical understanding necessary to effectively engineer these materials for devices ranging from solid oxide fuel cells to oxygen separation membranes and chemical looping applications. The interplay of material oxidation state and oxygen interstitial diffusion mechanism investigated here further advances the physical understanding of oxygen transport in this novel and emergent class of materials.

5. **Computational Methods**

All calculations were performed with Density Functional Theory (DFT) using the Vienna Ab Initio Simulation Package (VASP)[51, 52] code. The projector augmented wave method (PAW)[53] was used for the effective potential for all atoms. The PAW potentials used in these calculations have valence electron configurations of $2s^22p^4$ for O, $5s^25p^65d^16s^2$ for La, $3p^63d^94s^1$ for Ni and $4s^24p^65s^2$ for Sr. The generalized gradient approximation exchange-correlation functional of PW-91[54] was used with the Hubbard $U$ correction (GGA+$U$)[55, 56] applied to the Ni atoms with an effective $U$ value of 6.4 eV.[57] The stopping criteria for total energy calculations were 0.01 meV/cell and 0.1 meV/cell for the electronic and ionic relaxation, respectively. A plane wave cutoff energy of 425 eV was used in our calculations, consistent with previous DFT calculations on $La_{2-x}Sr_xNiO_{4+\delta}$ by Xie, et al.[20]

We have used two different supercell sizes of the same tetragonal (space group #138, $P4_2/ncm$) $La_{2-x}Sr_xNiO_{4+\delta}$ system with different Sr content $x$ and different oxygen interstitial content $\delta$: $2a \times 2a \times c$ (112 atoms) and $4a \times 4a \times c$ (448 atoms), where $a$ and $c$ are the lattice vectors parallel and perpendicular to the $La_{2-x}Sr_xO$ rocksalt layers, respectively of the 28-atom conventional cell of $La_2NiO_4$. As reported in the work of Li and Benedek,[26] the Ni-O octahedral rotation pattern of fully relaxed materials with $O_{int}$ defects ($\delta > 0$) corresponds to that of the tetragonal $P4_2/ncm$ phase. Hence, we used the tetragonal $P4_2/ncm$ phase for all the calculations in this work, which is consistent with the work of Li and Benedek.[26] The relaxed lattice parameters of pristine $La_2NiO_4$ material with zero strain are $a = 5.62$ Å and $c = 12.4$ Å.

When studying the effect of strain on oxygen interstitial diffusion, the epitaxial strain was applied to the *a-b* plane of the supercell, and the *c* direction is allowed to fully relax. The pristine supercells have 112 atoms ($La_{32}Ni_{16}O_{64}$) and 448 atoms ($La_{128}Ni_{64}O_{256}$), respectively. The corresponding oxygen interstitial content $\delta$ in these two supercells (single oxygen interstitial atom intercalation in the supercell) are: $\delta = 0.0625$ (112-atom supercell), $\delta = 0.0156$ (448-atom supercell). To find the transition state of each migration pathway and to calculate the corresponding migration energy barrier, we have used the climbing-image nudged elastic band (CI-NEB) method implemented in VASP.[58] Three images were used for all CI-NEB calculations; the sufficiency of using three images has been shown in the work of Li and Benedek.[26] Monkhorst-Pack[59] *k*-point meshes of $2 \times 2 \times 2$ and $1 \times 1 \times 1$ are used to sample the Brillouin zone for the $2a \times 2a \times c$ and $4a \times 4a \times c$ supercells, respectively, to achieve migration energy barriers converged within 10 meV/(oxygen interstitial). Based on the pristine supercells, the defected supercells were created by adding a single oxygen interstitial in the rocksalt layer. There are two possible defect states of the oxygen interstitial, the oxide state ($O_{int}^{2-}$) and the peroxide state ($O_{int}^{1-}$), which are visualized in **Figure 1** and **Figure 2** and discussed in **Section 2**. The calculation files showing the exact atomistic configurations of the pristine and defected supercells are provided as part of the **SI**.

In this work, we used the same La/Sr ordering and magnetic ordering (ferromagnetic) which were reported in the work of Xie, et al.[20] For the La/Sr ordering, the work of Xie, et al. used a set of special quasirandom structures (SQS) and predicted a lowest-energy configuration where all Sr atoms were located within one of the two rocksalt layers in the $2a \times 2a \times c$ supercell when Sr content $x \leq 0.5$.[20] Since the Sr content range explored in this work is $0 \leq x \leq 0.5$ (Ni valence of $Ni^{2+} \rightarrow Ni^{2.5+}$), the La/Sr ordering employed here is consistent with that of Xie, et al.[20] The oxygen interstitial is inserted into the rocksalt layer with no Sr atoms, which is the lowest-energy intercalation position of the oxide state ($O_{int}^{2-}$).[20] As shown in **Section 2**, $O_{int}^{2-}$ is less stable than $O_{int}^{1-}$ only when $x$ approaches 0.5 ($Ni^{2.5+}$), suggesting that $O_{int}^{2-}$ is the lowest-energy interstitial state over most relevant Sr concentrations. This higher stability of $O_{int}^{2-}$ relative to $O_{int}^{1-}$ over most Sr concentrations is the reason we used the rocksalt layer with no Sr atoms as the representative intercalation position for $O_{int}^{2-}$ and $O_{int}^{1-}$. For the specific case of $x = 0.5$, we tested the case of oxygen interstitial intercalation in the rocksalt layer with Sr atoms and found that our key conclusion about the emergence of the oxide-peroxide diffusion mechanism as competitive with the oxide-oxide mechanism around $x \sim 0.4$ ($\sim Ni^{2.4+}$) does not change with respect to varying the oxygen interstitial intercalation position. Additional details on this piece of analysis can be found in **SI Section 2**. Here we note that, due to the uncertainties in the impact of Sr ordering, it is difficult to carry out studies at conditions of higher Sr content than $x = 0.5$ (i.e., Ni valence > 2.5) and give quantitatively accurate results of interstitial formation energies and migration barriers. Therefore, in this work we limit the Sr content within the range of $x = 0$ to $x = 0.5$, and at the condition of high Sr content ($x = 0.4$ to $x = 0.5$), we cannot accurately predict whether the oxide-oxide mechanism or the oxide-peroxide mechanism should be the dominant diffusion mechanism. However, our results provide qualitative evidence that the oxide-peroxide mechanism emerges as a competitive diffusion pathway for oxygen interstitials at about $x \sim 0.4$ ($\sim Ni^{2.4+}$).

In the previous experimental work of Nakamura, et al.,[38] it was found that the oxygen interstitial content $\delta$ is very small at a relatively high Sr concentration, e.g. $\delta = 0.028$ at $x = 0.2$, $\delta = 0.003$ at $x = 0.3$, $\delta < 0.001$ at $x = 0.4$ ($T = 873$ K, $p(O_2) = 0.1$ atm, close to our modeling condition). Therefore, our calculations are based on an assumption where the oxygen interstitial concentration is in the dilute limit ($\delta \rightarrow 0$), and thus interactions between oxygen interstitial atoms are considered negligible. As the oxygen interstitial content $\delta$ becomes extremely low as the Sr content $x$ increases, the oxygen interstitial concentration may be sufficiently small that oxygen diffusion is no longer interstitial- or interstitialcy-mediated, but oxygen vacancy-mediated instead.[38] As we are interested in examining the Sr doping regime where oxygen transport is the result of interstitial motion, the range of Sr content variation studied in this work is from $x = 0$ to $x = 0.5$ ($Ni^{2.0+} \rightarrow Ni^{2.5+}$).

The four essential physical properties calculated in this study are the formation energies of the oxide and peroxide interstitials, $E_f(O_{int}^{2-})$ and $E_f(O_{int}^{1-})$, and the migration barriers of the oxide-oxide ($E_b^{oxide-oxide}$) and oxide-peroxide ($E_b^{oxide-peroxide}$) interstitial diffusion mechanisms as defined in **Section 2.1**. The equation for $E_f$ is: $E_f(O_{int}) = E(RP+O_{int}) - E(RP) - \mu_O$, where $E(RP+O_{int})$ is the DFT-calculated energy of the $La_{2-x}Sr_xNiO_{4+\delta}$ Ruddlesden-Popper material with one oxygen interstitial atom (also denoted as $E(O_{int}^{2-})$ or $E(O_{int}^{1-})$ for simplicity in the main text), $E(RP)$ is the DFT-calculated energy of the undefected $La_{2-x}Sr_xNiO_4$ Ruddlesden-Popper material, and $\mu_O$ is the chemical potential of oxygen, which is evaluated at $T = 773$ K and $p(O_2) = 0.2$ atm, the conditions of which are typical for devices containing materials that transport oxygen. Following previous studies,[60-64] the value of $\mu_O$ was shifted to account for the $O_2$ binding energy error in DFT, the finite temperature was introduced using data from the NIST-JANAF tables,[65] and the pressure component takes the form of a typical logarithmic ideal gas pressure shift to the chemical potential. Additionally, we assumed that the vibrational entropy contribution to the free energies of the $La_{2-x}Sr_xNiO_{4+\delta}$ solid ions on both sides of the defect formation reaction cancel,[60-64] while the vibrational free energy difference between oxygen in the gas phase and an oxygen interstitial in $La_{2-x}Sr_xNiO_{4+\delta}$ is incorporated into $\mu_O$. To do this, we have calculated the Einstein temperatures from vibrational frequencies of both $O_{int}^{2-}$ and $O_{int}^{1-}$ in $La_{2-x}Sr_xNiO_{4+\delta}$ using the finite-differences method in DFT. For $O_{int}^{2-}$, the Einstein temperatures were 583, 575 and 539 K, and for $O_{int}^{1-}$, the Einstein temperatures were 896, 597 and 461 K. Following previous studies,[61, 64] these Einstein temperatures were used in an Einstein model of the O vibrations to shift the value of $\mu_O$.

Here, we have summarized the precise methods used to calculate the previously discussed formation energies and migration barriers. These methods were used to simultaneously generate the most physically accurate results possible (additional details of any relevant tests or analysis are provided in the **SI**) while also minimizing the required calculation time, thus keeping the number and time of the calculations tractable. The changes of formation energies and migration barriers due to epitaxial strain were computed with 112-atom supercells to demonstrate qualitative trends.

$E_f(O_{int}^{1-})$: We used the 112-atom supercell with one oxygen interstitial atom for $E_f(O_{int}^{1-})$, which was found to be reasonable due to the negligible dependence of $E_f(O_{int}^{1-})$ on oxygen interstitial content. The calculated $E_f(O_{int}^{1-})$ values for both the 112-atom supercell ($\delta = 0.0625$) and 448-atom supercell ($\delta = 0.0156$) indicate a convergence within 0.01 eV (see **SI Section 1** for additional details).

$E_f(O_{int}^{2-})$: We used both the 112-atom ($\delta = 0.0625$) and 448-atom ($\delta = 0.0156$) supercells to calculate $E_f(O_{int}^{2-})$. Both cells were needed because $E_f(O_{int}^{2-})$ has a strong dependence on oxygen interstitial content. Knowing $E_f(O_{int}^{2-})$ for two concentrations, we then performed a linear extrapolation to obtain the dilute $E_f(O_{int}^{2-})$ at $\delta = 0$, assuming the relationship of $O_{int}^{2-}$ defect interaction with respect to the concentration $\delta$ is linear (data are explicitly shown in **SI Section 4**, **Table S1**).

$E_b^{oxide-oxide}$: We used the 448-atom supercell with one oxygen interstitial atom ($\delta = 0.0156$) and calculated the oxide-oxide diffusion barrier value. We used this larger 448-atom supercell instead of the 112-atom supercell to minimize spurious interactions between migrating O atoms in the periodic images.

$E_b^{oxide-peroxide}$: Through a series of preliminary calculations, we found that a linear relationship exists between $E_b^{oxide-peroxide}$ and $|E(O_{int}^{2-})-E(O_{int}^{1-})|$. The fitted linear function is $E_b^{oxide-peroxide} = 0.831 \times |E(O_{int}^{2-})-E(O_{int}^{1-})| + 0.518$ (unit eV for all terms), $R^2 = 0.988$. Additional details of this relationship are presented in **SI Section 3**. This scaling relationship was useful because it minimized the number of migration barrier calculations required, which for $E_b^{oxide-peroxide}$ would generally involve searching for the transition state via multiple different configurations with large 448-atom supercells. We calculated $E_b^{oxide-peroxide}$ for all Ni valence states using 112-atom supercells, and for only $Ni^{2.5+}$ using a 448-atom supercell. Then, this scaling relationship was used to obtain $E_b^{oxide-peroxide}$ for all Ni valences and all strain states using the $E(O_{int}^{2-})$ and $E(O_{int}^{1-})$ values from the 448-atom supercell defect energy calculations.

It should be noted here that the focus of this study is a qualitative discussion of the oxygen interstitial diffusion mechanisms with respect to Ni valence (Ni valence = $2+x+2\delta \approx 2+x$ ($\delta \to 0$), where $x$ is the Sr content) and strain. Quantitative effects of La/Sr ordering, magnetic structure, Ni-O octahedron rotation patterns at different temperatures and interactions between oxygen interstitial atoms on the results of oxygen interstitial defect formation energies and migration barriers are beyond the scope of this work. Therefore, our calculations only indicate qualitative trends of formation and migration energies with respect to Ni valence state and strain. We believe that the approximations to the above complicated factors used in this work are adequate to support the discovery and presented discussion of the oxide-peroxide diffusion mechanism in $La_{2-x}Sr_xNiO_{4+\delta}$. Given these approximations, we believe that additional focused studies to assess the impact of Sr ordering, interactions between oxygen oxide and peroxide interstitial configurations, and Ruddlesden-Popper magnetic and structural transitions would be highly beneficial to gain a more quantitative understanding of the defect chemistry and oxygen transport in this class of materials.

**Supporting Information**

Dependence of $E_f(O_{int}^{1-})$ on oxygen interstitial content $\delta$, migration barrier calculation in the (La,Sr)O rocksalt layer, linear scaling relation between $E_b^{oxide\text{-}peroxide}$ and $|E(O_{int}^{2-})-E(O_{int}^{1-})|$, tabulated data of all formation and migration energies in this work, and the key simulation input and output files for all DFT calculations performed in this work are available in the Supporting Information. In addition, a spreadsheet containing the numerical data required to make each data figure is also included.

**Conflicts of Interest**

The authors declare no competing financial interest.

**Acknowledgements**

The authors gratefully acknowledge funding from the National Science Foundation Software Infrastructure for Sustained Innovation (SI2) award No. 1148011. The research was also performed using the computer resources and assistance of the UW-Madison

Center for High Throughput Computing (CHTC) in the Department of Computer Sciences.

# Supporting Information

## "Factors controlling oxygen interstitial diffusion in the Ruddlesden-Popper oxide La$_{2-x}$Sr$_x$NiO$_{4+\delta}$"

Shenzhen Xu[1], Ryan Jacobs[1], Dane Morgan[1]

[1]*Department of Materials Science and Engineering, University of Wisconsin – Madison, Madison, WI, 53706, USA*

## S1. Dependence of $E_f(O_{int}^{1-})$ on oxygen interstitial content $\delta$

We have calculated $E_f(O_{int}^{1-})$ for both the 112-atom supercell ($\delta = 0.0625$) and 448-atom supercell ($\delta = 0.0156$) to test the dependence of $E_f(O_{int}^{1-})$ on oxygen interstitial content $\delta$. The specific testing case chosen had a composition of $La_{1.5}Sr_{0.5}NiO_{4+\delta}$ (Ni valence = 2.5) and the formation energies were: $E_f(\delta = 0.0625) = 1.03$ eV, $E_f(\delta = 0.0156) = 1.04$ eV. Thus, we can see that $E_f(O_{int}^{1-})$ is almost unchanged as $\delta$ approaches the dilute limit. Therefore, we can directly use the 112-atom supercell ($\delta = 0.0625$) for calculation of $E_f(O_{int}^{1-})$. As discussed in **Section 2.2** of the main text, a potential impact of adding oxygen interstitials is the Ni valence (oxidation state) of the system may change. The increase in Ni valence is significant for $O_{int}^{2-}$, where a higher concentration of $O_{int}^{2-}$ results in more oxidation. Therefore, when calculating $E_f(O_{int}^{2-})$ we explicitly use both the 112-atom and 448-atom supercells and extrapolate to the dilute limit as discussed in **Section 5** of the main text. However, as $O_{int}^{1-}$ is bonded to a lattice oxygen atom and forms an $O_2^{2-}$ dumbbell pair, the $O_{int}^{1-}$ has a negligible impact on Ni valence, thus explaining why $E_f(O_{int}^{1-})$ is nearly independent of $\delta$.

## S2. Migration barrier calculation in the (La,Sr)O rocksalt layer

The migration barriers $E_b^{oxide\text{-}oxide}$ and $E_b^{oxide\text{-}peroxide}$ presented in the main text were calculated by intercalating the $O_{int}^{2-}$ or $O_{int}^{1-}$ into the pure LaO rocksalt layer as discussed in **Section 5** of the main text. For $La_{2-x}Sr_xNiO_{4+\delta}$ (x > 0), an alternative oxygen intercalation layer is the (La, Sr)O mixed rocksalt layer. $O_{int}^{2-}$ ($O_{int}^{1-}$) is more stable in the pure LaO layer (mixed (La, Sr)O layer) based on our calculations and the work of Xie, et al.[1] This difference in defect stability in different rocksalt layers is the result of the following: (1) The $O_{int}^{2-}$ species dominates in terms of stability for most of the Ni valence range studied in this work (from about 2 to 2.35) based on our calculations in the pure LaO layer; (2) our test calculation at Sr content $x = 0.5$ (the highest Sr-content case in this work) showed that $E_f(O_{int}^{1-})$ is only lowered by 0.1 eV when its intercalation position changes from the LaO to the (La, Sr)O layer, suggesting that $O_{int}^{1-}$ stability is almost invariant with respect to different intercalation layers, and that the $O_{int}^{2-}$ species in the LaO-layer remains the most stable intercalation site in the Ni valence range from about 2.0 to 2.35. Therefore, we have chosen the pure LaO layer as the representative layer of oxygen intercalation for producing consistent results of oxygen interstitial diffusion and keeping the number of required calculations tractable.

Another issue related to oxygen interstitial placement in the LaO versus (La,Sr)O rocksalt layers is that $O_{int}^{1-}$ becomes more stable than $O_{int}^{2-}$ at $Ni^{2.5+}$ and $E_f(O_{int}^{1-})$ is lower in the (La,Sr)O mixed rocksalt layer than the pure LaO layer, therefore we have explicitly tested whether moving $O_{int}^{1-}$ to the (La,Sr)O mixed layer will affect our main conclusions about the existence of the oxide-peroxide diffusion mechanism at a Ni valence of about 2.5. These test calculations were carried out using the 112-atom supercell. At a Ni valence of 2.5 and no strain, $E_a^{oxide-peroxide}$ is 0.19 eV lower than $E_a^{oxide-oxide}$ when the oxygen interstitial is intercalated into the pure LaO layer. If we move the oxygen interstitial to the (La,Sr)O mixed layer and calculate $E_f(O_{int}^{2-})$, $E_f(O_{int}^{1-})$, $E_b^{oxide-oxide}$ and $E_b^{oxide-peroxide}$, the results are: $E_f(O_{int}^{2-}) = 1.74$ eV, $E_f(O_{int}^{1-}) = 0.92$ eV, $E_b^{oxide-oxide} = 0.90$ eV, $E_b^{oxide-peroxide} = 1.20$ eV. Thus, $E_a^{oxide-peroxide}$ is 0.52 eV lower than $E_a^{oxide-oxide}$, again indicating the oxide-peroxide mechanism is competitive with the oxide-oxide diffusion mechanism at a Ni valence of 2.5, which is qualitatively consistent with the results obtained when oxygen is intercalated into the pure LaO layer. Therefore, we conclude that having the oxygen intercalation position in the LaO versus the (La,Sr)O layer will not change our qualitative conclusion of the oxide-peroxide mechanism emerging as a competitive diffusion mechanism at high Ni valence values.

## S3. Linear scaling relation between $E_b^{oxide-peroxide}$ and $|E(O_{int}^{2-})-E(O_{int}^{1-})|$

In this section, we present the linear relationship between $E_b^{oxide-peroxide}$ and the formation energy difference $|E(O_{int}^{2-})-E(O_{int}^{1-})|$, shown in **Figure S1**. The data in **Figure S1** include $E_b^{oxide-peroxide}$ calculated in the 112-atom supercell with Ni valence = 2.0, 2.125, 2.25, 2.375, 2.5 under no strain, $E_b^{oxide-peroxide}$ calculated in the 112-atom supercell with Ni valence = 2.0, 2.25, 2.5 under 2% epitaxial tensile and 2% compressive strain and $E_b^{oxide-peroxide}$ calculated in the 448-atom supercell with Ni valence = 2.5 under no strain. From **Figure S1**, the data of $E_b^{oxide-peroxide}$ versus $|E(O_{int}^{2-})-E(O_{int}^{1-})|$ shows a strong linear relationship with $R^2 = 0.988$. The largest deviation between the fitted line and the data points is only about 100 meV.

This linear relationship makes physical sense based on the symmetrical diffusion pathway of the oxide-peroxide mechanism and the underlying principle of the Brønsted-Evans-Polanyi relations, which stipulate that the transition state energies of surface chemical reactions scale linearly with the difference in final and initial state energies. The Brønsted-Evans-Polanyi relations have been successfully used to explain the linear trend of transition state energies with reaction energies for both metal and transition metal oxide systems.[2,3] Here, larger values of $|E(O_{int}^{2-})-E(O_{int}^{1-})|$ correspond to a higher energy transition state, creating an overall larger migration barrier. Thus, a straightforward way to minimize $E_b^{oxide-peroxide}$ is to make $|E(O_{int}^{2-})-E(O_{int}^{1-})| = 0$, that is, to make the $O_{int}^{2-}$ and $O_{int}^{1-}$ degenerate in energy.

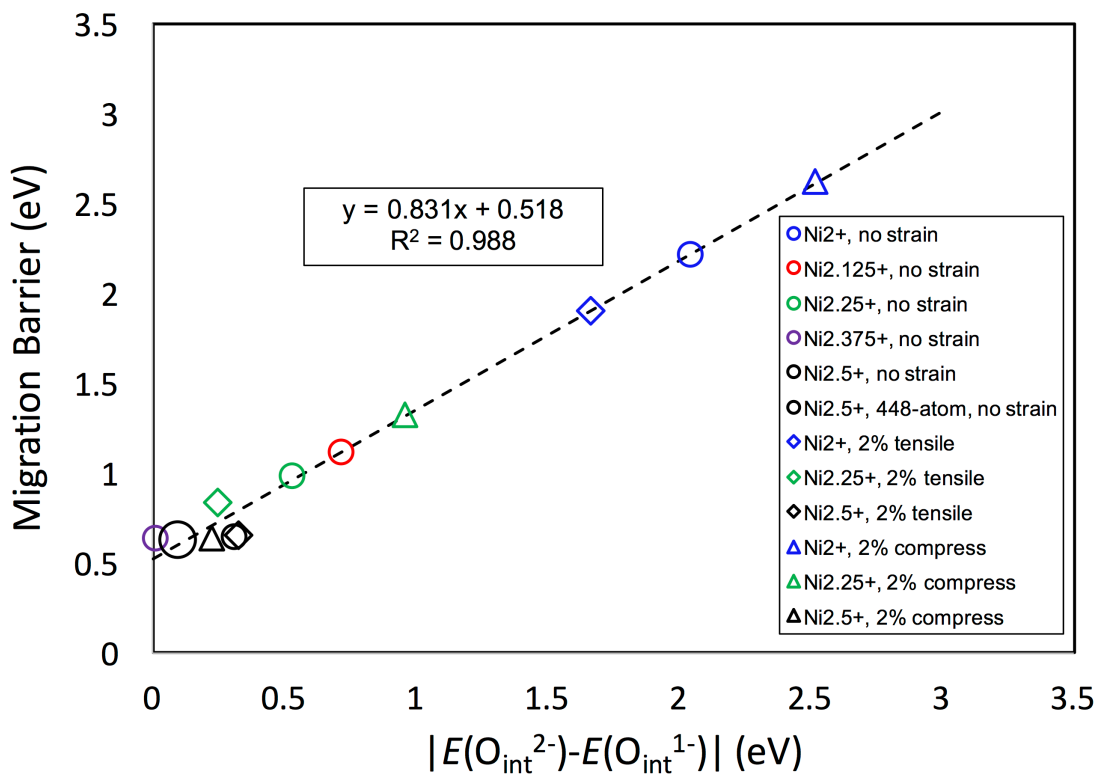

**Figure S1.** Linear relationship between $E_b^{oxide-peroxide}$ and $|E(O_{int}^{2-})-E(O_{int}^{1-})|$. The blue, red, green, purple, and black data points indicate Ni valences of 2, 2.125, 2.25, 2.375, and 2.5, respectively. The circle, diamond, and triangle symbols indicate no strain, 2% tensile strain, and 2% compressive strain, respectively. All data points were obtained for 112-atom supercells, except for the large black circle data point, which was obtained from a 448-atom supercell. The dashed line is the best fit line.

## S4. Tabulated data of all formation and migration energies in this work

**Table S1.** Formation and migration energies under no strain. Units of all quantities are (eV).

| Ni valence | $E_f(O_{int}^{1-})$ 112-atom | $E_f(O_{int}^{2-})$ 112-atom | $E_f(O_{int}^{2-})$ 448-atom | $E_f(O_{int}^{2-})$ ($\delta \to 0$) | $E_b^{oxide-peroxide}$ (linear scaling) | $E_b^{oxide-oxide}$ 448-atom |
|---|---|---|---|---|---|---|
| 2 | 0.79 | -1.26 | -1.72 | -1.88 | 2.73 | 0.31 |
| 2.125 | 0.95 | 0.23 | -0.51 | -0.75 | 1.93 | 0.33 |
| 2.25 | 1.01 | 0.48 | 0.31 | 0.26 | 1.14 | 0.39 |
| 2.375 | 0.97 | 1.96 | 0.98 | 1.00 | 0.54 | 0.46 |
| 2.5 | 1.03 | 1.33 | 1.13 | 1.07 | 0.55 | 0.70 |

**Table S2.** Formation energies under 2% epitaxial tensile and compressive strain, calculated with the 112-atom supercell. Units of all quantities are (eV).

| Ni valence | $E_f(O_{int}^{1-})$ no strain | $E_f(O_{int}^{1-})$ 2% tensile | $E_f(O_{int}^{1-})$ 2% compressive | $E_f(O_{int}^{2-})$ no strain | $E_f(O_{int}^{2-})$ 2% tensile | $E_f(O_{int}^{2-})$ 2% compressive |
|---|---|---|---|---|---|---|
| 2 | 0.79 | 0.76 | 0.95 | -1.88 | -1.53 | -2.19 |
| 2.25 | 1.01 | 0.92 | 1.15 | 0.26 | 0.44 | -0.04 |
| 2.5 | 1.03 | 0.92 | 1.20 | 1.07 | 0.97 | 1.15 |

**Tables S3.** Migration barriers under 2% epitaxial tensile and compressive strain, calculated with the 112-atom supercell. Units of all quantities are (eV).

| Ni valence | $E_b^{oxide-peroxide}$ no strain | $E_b^{oxide-peroxide}$ 2% tensile | $E_b^{oxide-peroxide}$ 2% compressive | $E_b^{oxide-oxide}$ no strain | $E_b^{oxide-oxide}$ 2% tensile | $E_b^{oxide-oxide}$ 2% compressive |
|---|---|---|---|---|---|---|
| 2 | 2.21 | 1.90 | 2.62 | 0.36 | 0.29 | 0.44 |
| 2.25 | 0.98 | 0.83 | 1.32 | 0.44 | 0.36 | 0.52 |
| 2.5 | 0.65 | 0.65 | 0.64 | 0.51 | 0.45 | 0.59 |